\documentclass[12pt]{iopart}

\usepackage{bm}
\usepackage{epsfig}
\usepackage{citesort}
\usepackage{graphicx}
\eqnobysec

\newcommand{\lwig}{\mbox{\;\raisebox{.3ex}
    {$<$}$\!\!\!\!\!$\raisebox{-.9ex}{$\sim$}\;}}
\newcommand{\gwig}{\mbox{\;\raisebox{.3ex}
    {$>$}$\!\!\!\!\!$\raisebox{-.9ex}{$\sim$}}\;}
\newcommand{\lambdabar}{{\hbox{$\lambda$\kern-1.ex\raise+0.45ex\hbox{--}}}}

\begin{document}

\begin{flushright}
{\large \tt MPP-2006-15}
\end{flushright}

\title{Measuring neutrino masses and dark energy with weak lensing tomography}

\author{Steen~Hannestad}
\address{Department of Physics and Astronomy \\
University of Aarhus, DK-8000 Aarhus C, Denmark}

\author{Huitzu~Tu}
\address{Department of Physics and Astronomy \\
University of Aarhus, DK-8000 Aarhus C, Denmark}

\author{Yvonne~Y.~Y.~Wong}
\address{Max-Planck-Institut f\"ur Physik (Werner-Heisenberg-Institut) \\
F\"ohringer Ring 6, D-80805 M\"unchen, Germany}

\ead{\mailto{sth@phys.au.dk},
\mailto{huitzu@phys.au.dk}, \mailto{ywong@mppmu.mpg.de}}

\begin{abstract}
Surveys of weak gravitational lensing of distant galaxies will be
one of the key cosmological probes in the future. We study the
ability of such surveys to constrain neutrino masses and the equation of
state parameter of the dark energy, focussing on how tomographic
information can improve the
sensitivity to these parameters.  We also provide a detailed
discussion of systematic effects pertinent to weak lensing
surveys, and the possible degradation of sensitivity to cosmological
parameters due to these effects.  For future probes such as the Large Synoptic
Survey Telescope survey, we find that, when combined with cosmic
microwave background data from the Planck satellite, a sensitivity
to neutrino masses of $\sigma(\sum m_\nu) \lwig 0.05$ eV can be
reached.  This results is robust against variations in the running of the
scalar spectral index, the time-dependence of dark energy equation of state,
and/or the number of relativistic degrees of freedom.

\end{abstract}
\maketitle

\section{Introduction}

With the advent of precision measurements of the cosmic microwave
background (CMB), large scale structure (LSS) of galaxies, and
distant type Ia supernovae (SNIa), a new paradigm of cosmology has
been established.  In this new standard model, the geometry is flat so
that $\Omega_{\rm total} = 1$, and the total energy density is made up
of matter ($\Omega_m \sim 0.3$) [comprising of baryons ($\Omega_b \sim 0.05$)
and cold dark matter ($\Omega_{\rm CDM} \sim 0.25$)],
and dark energy ($\Omega_{\rm de} \sim 0.7$, with equation of state
$w \equiv P/\rho\simeq -1$).
With only a few free parameters this
model provides an excellent fit to all current
observations~\cite{Riess:1998cb,Perlmutter:1998np,Spergel:2003cb,bib:sdss1}.
In turn, this allows for constraints on other, nonstandard
cosmological parameters.

Two very interesting possibilities discussed widely in the literature
are: (i) a subdominant contribution to the total energy density in the form of
neutrino hot dark matter (HDM) and hence limits on the neutrino mass $m_{\nu}$,
and (ii) an equation of state for the dark
energy component that deviates from $w=-1$ and/or is time-dependent.

In the first case, data from atmospheric and solar neutrino oscillation
experiments strongly suggest two important mass differences in the neutrino
mass hierarchy, $\Delta m^2_{\rm atm} \simeq |\Delta m^2_{23}|
\simeq  2.4 \times 10^{-3} \ {\rm eV}^2$ and
$\Delta m^2_{\rm sun} \simeq \Delta m^2_{12} \simeq 7.9 \times 10^{-5} \ {\rm eV}^2$
(e.g.,~\cite{Fogli:2005cq}).
The simplest interpretation of this data set implies $m_1 \sim 0$,
$m_2 \sim \sqrt{\Delta m^2_{\rm sun}}$ and $m_3 \sim  \sqrt{\Delta m^2_{\rm atm}}$, i.e.,
the so-called normal hierarchy.  Alternatively,  since the sign of $\Delta m^2_{23}$ cannot
be fixed by atmospheric neutrino oscillations,
an inverted hierarchy with
$m_3 \sim 0$, $m_2 \sim \sqrt{\Delta m^2_{\rm atm}}$ and $m_1 \sim  \sqrt{\Delta m^2_{\rm atm}}$
is also allowed.  Yet a third possibility is that the three mass eigenstates are nearly
degenerate, with $m_1 \sim m_2 \sim m_3 \gg \sqrt{\Delta m^2_{\rm atm}}$.  If the last case prevails,
then oscillation experiments can offer no useful information on the absolute neutrino
mass scale.

Cosmology provides an interesting measure of the neutrino mass
through the neutrino's kinematic effects on large scale structure
formation (e.g., \cite{Hannestad:2006zg}). Qualitatively, free
streaming of a HDM particle causes  essentially all of its own
perturbations to be erased on scales below its free streaming
length $\lambda_{\rm fs}$. This erasure is simultaneously
propagated to other species (eg., CDM) via the metric source term.
The net effect on the present day large scale matter power
spectrum $P_m(k)$ is a suppression of order $\Delta
P_{m}(k)/P_m(k) \sim -8 \ \Omega_{\nu}/\Omega_m$ (assuming
$\Omega_{\nu} \ll \Omega_m$, and normalisation at $k\to0$) at
large wavenumbers $k \gg k_{\rm fs} \sim 2 \pi/\lambda_{\rm fs}$,
where $\Omega_\nu h^2 = \sum m_{\nu}/(92.5 \ {\rm eV})$.  Indeed,
within the $\Lambda$CDM framework, LSS observations to date have
already been able to constrain the sum of the neutrino mass  to
$\sum m_\nu \lwig 0.3 \to 1 \ {\rm eV}$ ($2 \sigma$)
\cite{bib:hannestad2003,Elgaroy:2004rc,Barger:2003vs,Crotty:2004gm,Seljak:2004xh,%
Fogli:2004as,Tegmark:2005cy,Hannestad:2003ye,Goobar:2006xz},
depending on the data set used. This limit should be compared with
the current laboratory bound from tritium $\beta$ decay
experiments, $m_{\nu} < 2.2 \ {\rm eV}$ ($2 \sigma$)
\cite{bib:mainz,bib:troitsk}, and the projected sensitivity of the
upcoming KATRIN experiment, $\sim 0.2 \ {\rm eV}$
\cite{bib:katrin,katrin2}.

In the second case, the simplest candidate for dark energy is a
cosmological constant~$\Lambda$, obeying $w=-1$.
A minimal extension of this is to assume that the dark energy equation of
state is given by $P=w \rho$, where $w$ is constant in time.
Using this framework, various
cosmological data sets have been analysed in many independent
studies (e.g., \cite{Bean:2001xy,Hannestad:2002ur,Melchiorri:2002ux}).
With the most recent data $w$ is constrained to be
extremely close to $-1$, the value for a pure cosmological constant
(e.g., \cite{snls}).

However, in many particle physics models of dark energy, the
equation of state is expected to depend on the scale factor,
and many authors have analysed the data in terms of dark
energy models with varying equations of state
(e.g., \cite{Hannestad:2004cb,Wang:2004py,Upadhye:2004hh}). From an
observational point of view, the simplest possibility is that $w$
undergoes a transition between two asymptotic values. This is very
often also the case in particle physics models of dark energy such as
quintessence \cite{Wetterich:1987fm,Ratra:1987rm,Peebles:1987ek}.
The same is also true in models where dark energy
is a consequence of modified gravity \cite{Deffayet:2001pu,Dvali:2003rk}.
Cosmological data to date show no evidence for time variations in the
dark energy equation of state.  However, for future
high precision observations, it is natural to search for such time
dependences.  Later on when we analyse the ability of future cosmological
probes to constrain neutrino masses and the dark energy equation of state,
we will also include this possibility using a simple parameterisation.

Recently, it was shown that presently available data admit an
apparent degeneracy between $m_\nu$ and $w$
\cite{Hannestad:2005gj}. Specifically, by extending $w$ (taken to
be constant in time) into the phantom regime ($w < -1$), the
neutrino mass bound relaxes to $\sum m_{\nu} \leq 1.48 \ {\rm eV}$
from a combined CMB+LSS+SNIa analysis, compared to $\sum m_\nu
\leq 0.65 \ {\rm eV}$ for the case of a fixed $w = -1$ using the
same data set. This $m_{\nu},w$ degeneracy can be explained as
follows.  The suppression of power at small scales due to a large
$m_{\nu}$ can be accommodated by LSS data if $\Omega_m$ is
simultaneously increased to ``restore'' the matter power spectrum
to its canonical spectral shape. On the other hand, $w$ and
$\Omega_{\rm de}$ (and hence $\Omega_m = 1 - \Omega_{\rm de}$ in a
flat universe) are degenerate parameters in the Hubble expansion
rate  at redshifts probed by SNIa, such that an increased
$\Omega_m$ can be compensated for by a decreased~$w$. These two
degeneracies then combine to produce an apparent interplay between
$m_{\nu}$~and~$w$.

It has been shown in \cite{Goobar:2006xz} that the $m_\nu,w$
degeneracy can be effectively broken by adding to the likelihood
analysis new data from the Sloan Digital Sky Survey (SDSS)
measurement of the baryon acoustic oscillation (BAO) peak
\cite{Eisenstein:2005su}, bringing the present bound on the sum of
neutrino masses to the $0.3 \to 0.5$ eV range. In this work,
however, we consider the prospects for breaking this degeneracy
with an alternative approach, namely, weak gravitational lensing.

Weak gravitational lensing (or cosmic shear) of distant galaxies by the intervening
large scale structure provides an unbiased way to
map the matter distribution in the universe
(e.g., \cite{Bartelmann:1999yn,VanWaerbeke:2003uq,Refregier:2003ct}).
Perturbations in the density field between the source
and the observer induce distortions in the images of source galaxies.   By measuring the
angular correlation of these distortions, one can in principle reconstruct
the underlying three-dimensional matter power spectrum, without making assumptions
about the luminous versus dark matter bias
(cf.\ conventional galaxy surveys).
Since its first detection in 2000
\cite{vanWaerbeke:2000rm,Kaiser:2000if,Bacon:2000sy,Wittman:2000tc},
the sensitivity of
weak lensing surveys has improved to the stage where they are able to provide
constraints on the matter density $\Omega_m$ and the fluctuation amplitude $\sigma_8$
that are competitive with other cosmological probes (e.g., \cite{Hoekstra:2005cs,Semboloni:2005ct}).
Several larger surveys are now in operation or in planning.
These include the Canada--France--Hawaii Telescope Legacy Survey (CFHTLS) \cite{bib:cfhtls},
the Dark Energy Survey (DES) \cite{Abbott:2005bi},
Pan-STARRS \cite{bib:panstarrs}, and VISTA
\cite{bib:vista}, culminating in the most ambitious SuperNova/Acceleration Probe (SNAP)
\cite{bib:snap} and
Large Synoptic Survey Telescope (LSST) \cite{bib:lsst} surveys.

Furthermore, all future weak lensing surveys will provide
photometric redshift information on the source galaxies.  This
additional information allows one to study tomographically the
growth of the intervening large scale structure and the
distance--redshift relation \cite{Hu:1999ek}. Indeed, weak lensing
tomography is expected to reach a level of sensitivity comparable
(and orthogonal) to  SNIa for dark energy parameter determination
(e.g.,~\cite{Huterer:2001yu,Hu:2002rm,Abazajian:2002ck,Ishak:2003zw,Song:2003gg,Takada:2003ef,Ishak:2005we}).

The structure of this paper is as follows.  We outline the principle of
weak lensing tomography and present the formalism for computing the convergence
power spectrum in section \ref{sec:tomography}.  In section \ref{sec:systematics},
we discuss some of the systematic effects relevant for tomographic
studies.  The Fisher matrix formalism and the projected errors
for various cosmological parameters
are presented in section \ref{sec:forecast}.
Section \ref{sec:conclusions} contains our conclusions.

\section{Weak lensing tomography \label{sec:tomography}}

Distortions in the images of galaxies induced by density perturbations between
the source and the observer can be described by the distortion tensor (e.g.,
\cite{Bartelmann:1999yn,Refregier:2003ct}),
\begin{equation}
\psi_{ij} \equiv \frac{\partial \theta_s^i}{\partial \theta^j} - \delta_{ij} =
\left( \begin{array}{cc}
         - \kappa - \gamma_1 & - \gamma_2 \\
        - \gamma_2 &  -\kappa + \gamma_1 \\
        \end{array} \right)\, ,
\end{equation}
where ${\bm \theta}_s$ and ${\bm \theta}$ denote the angular
position of a light ray in the source and the image planes
respectively, and the tensor $\psi_{ij}$
decomposes into a convergence
(i.e., magnification, $\kappa \ll 1$) and two shear
(i.e., stretching, $\gamma_1,\gamma_2 \ll 1$) components.
In the Born approximation, the convergence component
is given by a weighted projection of the density fluctuations
$\delta \equiv \delta \rho/\rho$ along the line
of sight,
\begin{equation}
\label{eq:convergence}
\kappa({\bm \theta}) = \int^{\chi_h}_0 d \chi \ W(\chi) \delta(\chi,\chi {\bm \theta})\, ,
\end{equation}
where $\chi  = \int^z_0 \  d z/H(z)$ is the comoving radial
distance, $H (z)$ the Hubble parameter, and $\chi_h$ denotes the distance to
the horizon.
The assumption of a flat universe is implicit.

The lensing weighting function is defined as
\begin{equation}
\label{eq:weightingfunction}
W(\chi) = \frac{W_0}{\bar{n}_{\rm gal}} a^{-1} (\chi) \chi
\int^{\chi_h}_{\chi} d \chi' n_{\rm gal}(\chi') \frac{\chi'-\chi}{\chi'}\, ,
\end{equation}
with $W_0=(3/2) \ \Omega_m H_0^2$, and $n_{\rm gal}(\chi) = H(z) \ n_{\rm gal}(z)$,
where $n_{\rm gal}(z)$ is the number of source galaxies between
redshift $z$ and $z + dz$ per steradian, and satisfies the normalisation condition
$\int^{\infty}_0 dz \ n_{\rm gal}(z) = \bar{n}_{\rm gal}$.  It is
generally assumed that $n_{\rm gal}(z)$ takes the form
\begin{equation}
\label{eq:ngal}
  n_{\rm gal}(z) = \bar{n}_{\rm gal} \ \frac{\beta}{z_0 \Gamma \left( \frac{1+\alpha}{\beta}\right)}
  \left( \frac{z}{z_0} \right)^{\alpha} \exp \left[ -
\left(\frac{z}{z_0}\right)^{\beta} \right]\, ,
\end{equation}
where the parameters $\alpha$, $\beta$ and $z_0$ are chosen to fit the
observed galaxy redshift distribution.
Here, we fix $\alpha=2$ and $\beta=2$.
The choice of the parameter $z_0$ depends on the type of weak lensing
survey: for a ground-based ``wide'' survey,
$z_0 = 1$ corresponds to a median galaxy redshift $z_{\rm med} \sim 1$; for
a space-based ``deep'' survey, the choice of $z_0 = 1.5$ gives $z_{\rm med} \sim 1.5$
(e.g., \cite{Massey:2003xd,Refregier:2003xe}).

Where photometric redshift information on the source galaxies is available and the
sources separable into tomography bins,
equations (\ref{eq:convergence}) and (\ref{eq:weightingfunction}) can be generalised
to give the convergence of the $i$th bin via
\begin{equation}
\kappa ({\bm \theta}) \to \kappa_i ({\bm \theta}),
\quad W(\chi) \to W_i (\chi), \quad n_{\rm gal}(z) \to n_i(z), \quad
 \bar{n}_{\rm gal} \to \bar{n}_i\, ,
\end{equation}
where $\bar{n}_i = \int^{\infty}_0 dz \ n_i(z)$ is the number density of source
galaxies in the $i$th tomography bin,
and $\sum_i n_i(z) = n_{\rm gal}(z)$ by construction.  A simple model for $n_i(z)$ would be
$n_i(z)=\Theta(z_{i+1}-z)  \Theta(z-z_i) n_{\rm gal}(z)$, where $z_i$ and $z_{i+1}$ denote
the lower and upper limits of the $i$th tomography bin.

\subsection{Lensing observables}

Since lensing by large scale density perturbations leads to image distortions typically
of order $1 \ \%$ or less (much smaller than the intrinsic scatter in galaxy shapes),
cosmological weak lensing must be measured statistically with a large
ensemble of sources and lenses.
By decomposing the convergence field into  two-dimensional Fourier modes,
\begin{equation}
 \kappa_i({\bm \theta}) = \sum_{\bm \ell} \tilde{\kappa}^i_{\bm \ell} \exp(i {\bm \ell}
 \cdot {\bm \theta})\, ,
\end{equation}
one can define the angular power spectrum of the convergence
at multipole $\ell$ between the $i$th and $j$th tomographic bin as
\begin{equation}
 \langle \tilde{\kappa}^i_{\bm \ell} \ \tilde{\kappa}^{j*}_{{\bm \ell}'} \rangle = (2 \pi)^2
 \delta({\bm \ell}-{\bm \ell}') C^{ij}_{\ell}\, .
\end{equation}
In the Limber approximation (where only modes with longitudinal wavenumbers
$k_3 \lwig  \chi^{-1} < \ell/\chi$ contribute to the correlation),
the convergence power spectrum is related to the three-dimensional
matter power spectrum $P_m(k)$ via \cite{Kaiser:1991qi,Kaiser:1996tp}
\begin{equation}
\label{eq:wlpower}
  C^{i j}_{\ell} = \int^{\chi_h}_0 d \chi \ W_i (\chi) \ W_j (\chi) \ \chi^{-2} \
  P_m \left( k=\ell/\chi, z \right)\, ,
\end{equation}
assuming a flat universe.  Thus
weak lensing tomography with $n_{\rm t}$ bins yields
$n_{\rm t}$ auto ($i=j$) and $n_{\rm t}(n_{\rm t}-1)/2$ cross ($i \neq j$) correlation spectra.
The case of $n_{\rm t}=1$ corresponds to lensing without tomography, and we shall be
using these two terms interchangeably throughout the text.

\begin{figure}[t]
\hspace*{25mm}
\epsfxsize=10cm \epsfbox{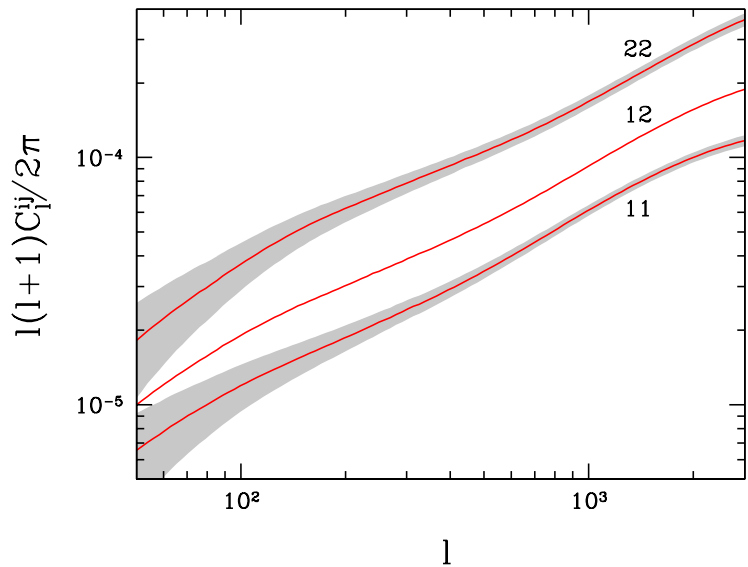}
\caption{\label{fig:cl} Convergence auto and cross power spectra for the
standard $\Lambda$CDM model, assuming the galaxy distribution function (\ref{eq:ngal})
and two tomography bins at $0 \leq z\leq 1.5$ and $1.5 < z \leq 3$.  The $1 \sigma$
error bands are computed for a lensing survey with sky coverage $f_{\rm sky}=0.01$,
$\bar{n}_{\rm gal}=30 \ {\rm arcmin}^{-2}$,
and $\gamma_{\rm rms}=0.4$, smeared over bands of
width $\Delta \ell = \ell/4$.
See section \ref{sec:forecast}
for definitions.}
\end{figure}

Figure~\ref{fig:cl} shows an example of the convergence auto and cross power
spectra expected for tomography with two bins.
It is important to note that these power spectra are strongly correlated,
and in their entirety do not provide as much information as a na\"{\i}ve interpretation
 might suggest.
The gain in information from tomographic binning can be quantified
by the correlation coefficient between the power spectra of the individual
tomography bins, $R^{i j} \equiv C^{i j}_\ell/\sqrt{C^{i i}_\ell\, C^{j j}_\ell}$
~\cite{Hu:1999ek}.  A strong correlation (i.e., $R^{i j} \sim 1$) between two bins
means that further subdivision is futile.

Note that while we have discussed only the convergence, existing
and future weak lensing surveys will measure the source galaxies'
ellipticities and hence the two shear components. However, by
decomposing the shear signal into a gradient (``$E$-mode'') and a
curl (``$B$-mode'') component, it can be shown that the statistics
of the $E$-mode shear and of the convergence are identical (e.g.,
\cite{Dodelson:2003ft}). On the other hand, the $B$-mode is
expected to yield a vanishing spectrum and therefore serves as a
powerful test for residual systematics
\cite{VanWaerbeke:2004af,Jarvis:2004js}. Henceforth, we shall
consider the convergence power spectrum as representative of the
lensing signatures.

\subsection{Dark energy and neutrinos\label{sec:denu}}

\begin{figure}[t]
\hspace*{25mm}
\epsfxsize=10cm \epsfbox{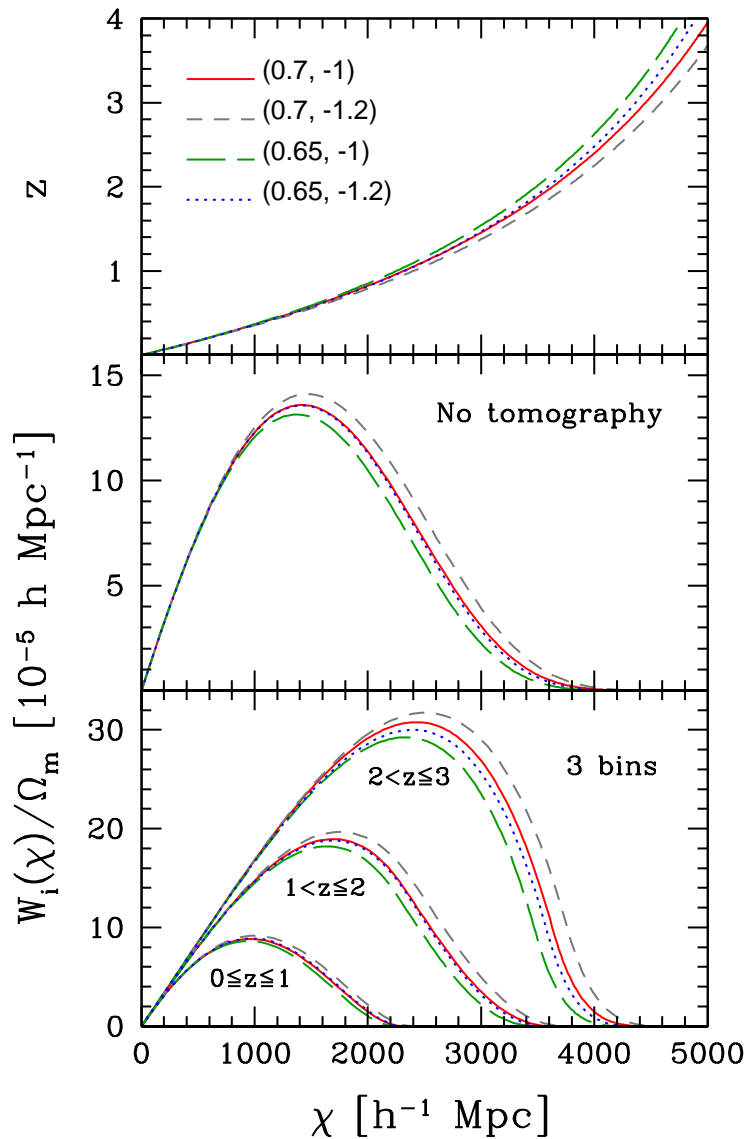}
\caption{{\it Top}: Distance--redshift relation for four sets of
dark energy parameters $\{\Omega_{\rm de},w\}$ listed in the Figure,
where $w$ is taken to be constant in time.
{\it Middle}:  Lensing weights $W(\chi)/\Omega_m$, without tomography,
for the same four sets of $\{\Omega_{\rm de},w\}$,
assuming the galaxy distribution function
(\ref{eq:ngal}) with $\alpha=2$, $\beta=2$, and $z_0=1$.
{\it Bottom}: Lensing weights for three tomographic divisions at
$0 \leq z \leq 1$, $1 < z \leq 2$, and $2 < z \leq 3$.
\label{fig:weights}}
\end{figure}

The  weak lensing signal (\ref{eq:wlpower}) comprises of
two distinct components: the lensing weights $W(\chi)$ and
the matter power spectrum $P_m(k)$.  While the latter quantifies the
universal matter distribution, the role of the former, in broad terms,
is to project this three-dimensional distribution onto the two-dimensional
sky for the observer.  In general, for a single source at comoving
distance $\chi_0$, the function $W(\chi)$  peaks at roughly
$\chi_0/2$.  Thus, within the $\Lambda$CDM framework,
a typical source distribution such as (\ref{eq:ngal})
peaking at $z \sim 1$ will be
most efficiently lensed by structures at $z \sim 0.5$ (cf.\ Figure \ref{fig:weights}).

The effects of dark energy are manifest in both the lensing weights and the matter
power spectrum.
In the case of $W(\chi)$,
the dark energy density $\Omega_{\rm de}$ and equation of state $w(z)$
determine the Hubble expansion rate at low redshifts,
\begin{equation}
\label{eq:hubbleparam}
 H^2 (z)= H_0^2 \ \left[(1-\Omega_{\rm de}) \ (1+z)^3 + \Omega_{\rm de} \
 e^{3 \times \int^z_0 d z^\prime\, [1+w(z^\prime)]} \right],
\end{equation}
and hence the distance--redshift relation, $\chi=\int^z_0
dz/H(z)$, responsible for the projection (cf.\ Figure
\ref{fig:weights}). In the above equation we have assumed spatial
flatness. This assumption will be kept throughout the analysis.

On the other hand, dark energy suppresses the growth of structure
relative to the case of a flat, matter-dominated universe.  In the
absence of massive neutrinos, the linear matter power spectrum
evolves  as $P_m(k,a)= D^2(a) P_m(k,a=1)$, independently of $k$.
Here, $D(a)$ is the growth function normalised to $D(a=1)=1$, and
can be obtained by solving the differential equation (e.g.,
\cite{Wang:1998gt})
\begin{equation}
2 \frac{d^2g}{d \ln a^2} + [5 - 3 w(a) \Omega_{\rm de} (a)] \frac{d g}{d \ln a}
+ 3 [1-w(a)] \Omega_{\rm de}(a) g(a)=0,
\end{equation}
where $g(a)=D(a)/a$, and $\Omega_{\rm de}(a)$ is the dark energy density
at epoch $a$.  The effects of $\Omega_{\rm de}$ and $w$ on the growth function are
generally mild compared to those on $W(\chi)$.  For the case of $w=-1.2$, for example,
the lensing weight $W(\chi)$ increases by a uniform $3 \to 4 \ \%$ relative to the $w=-1$
case across $0<z<2$, while the change in the growth function $D(a)$ is a minute $-0.05 \ \%$ at
$z=0.5$ at the peak of the lensing efficiency and only reaches $-2 \ \%$ at $z=2$ at
the tail of $W(\chi)$.
Thus the dominant constraints
on the dark energy parameters will come from projection.
In addition, $\Omega_{\rm de}$ ($=1-\Omega_m$) impacts immediately on the
epoch of matter--radiation equality, which in turn determines the location of the
turning point in the matter power spectrum.

Neutrino masses in the sub-eV range do not play a role in the universal
expansion rate at low redshifts.  Their early free-streaming
effect on the shape of the matter power spectrum is however well known.
As mentioned earlier, the loss of power at large wavenumbers
due to a large $m_{\nu}$ can be offset by simultaneously translating the matter
power spectrum towards large values of $k$.  This translation can be accomplished by
increasing (decreasing) $\Omega_m$ ($\Omega_{\rm de}$) and hence moving
the matter--radiation equality epoch forward to an earlier time when
 the causal horizon is smaller.
In this way, $m_{\nu}$ and $\Omega_{\rm de}$
form a degenerate pair.

Lensing without tomography, like angular galaxy surveys, makes use of a projected signal
accumulated over a large distance ($\sim 4000 \ h^{-1} {\rm Mpc}$).  The information
content of this signal, especially  in the radial $\chi$ direction,
is therefore limited.
Tomography binning allows one to recover some radial information through a set of
lensing weights $W_i(\chi)$ that peak at different values of $\chi$.
This is particularly useful for establishing the distance--redshift relation
and hence breaking the $\Omega_{\rm de},w$ degeneracy.  As shown in
Figure \ref{fig:weights}, two sets of $\{\Omega_{\rm de},w\}$ with
lensing weights that are indistinguishable to the naked eye in the case of no tomography
begin to show perceivable signs of differences
when three tomography bins are used.

Tomography will not improve directly the neutrino mass bound.
Even though massive neutrinos continue to affect the growth of small scale density
fluctuations at low redshifts, the shape of the matter power spectrum evolves by a
negligible amount over the $z$ range probed by weak lensing.  To illustrate this
point, we note that for $\sum m_\nu=0.3 \ {\rm eV}$,
the normalised power spectrum $P_m(k)/P_m(k \to 0)$ changes
by less than $2 \ \%$ at large values of $k$ between $z=0$ and $z=1.5$.
This is smaller than the overall power suppression
expected even for a
$0.07 \ {\rm eV}$ neutrino today
($\sim 8 \ \Omega_{\nu}/\Omega_m \sim 5 \ \%$).
Nonetheless, once the $\Omega_{\rm de},w$ degeneracy
is lifted by projection effects, the $\Omega_{\rm de},m_{\nu}$ shape degeneracy
will be correspondingly reduced and the bound on $\sum m_\nu$ tightened.

\section{Systematic effects \label{sec:systematics}}

A number of systematic effects have been identified that could severely limit the potential
of weak lensing surveys for cosmological parameter extraction.
We discuss some of these below.
For other systematics not covered here, we refer the reader to \cite{Refregier:2003ct}.

\subsection{Photometric redshift uncertainties}

Future surveys will observe a projected billion galaxies, a sample too large
for individual spectroscopic redshift measurements.
By observing the galaxy broad band features (i.e., broad band photometry)
and comparing them to
predictions from galaxy spectral energy distributions,
redshift information can be obtained much more quickly at
the cost of lower accuracy.

The uncertainties associated with the photometrically measured redshift are typically
 $\pm 0.034 \to 0.1$
(Hubble Deep Field~\cite{Fernandez-Soto:1998es}, CFHTLS
\cite{Semboloni:2005ct}, SNAP with IR-filters~\cite{Massey:2003xd}, Combo-17
survey~\cite{Brown:2002wt} with multi-colour filters etc.),
compared with $\pm 0.001$ using  spectroscopic methods (DEEP2
\cite{bib:deep2}, VVDS \cite{bib:vvds}, etc.). Assigning redshifts
to galaxies using photometric methods should not contribute
significantly to the error budget in weak lensing (without
tomography) measurements. For tomographic studies, however,
binning with photometric redshifts $z_{\rm ph}$ that suffer from
uncontrolled scatter and bias relative to the true (spectroscopic)
redshifts $z$ can lead to considerable error degradation in the
dark energy  parameter extraction. The authors of
\cite{Huterer:2005ez,Ma:2005rc} found that the scatter and the
bias per redshift interval $\delta z=0.1$ need to be known to an
accuracy of $0.003 \to 0.01$, in order not to degrade the errors in
$\Omega_{\rm de}$ and $w$ by more than a factor of~$1.5$.

We account for uncertainties in the photometric redshift
following the approach of \cite{Ma:2005rc}.
The true number of galaxy per steradian at redshift $[z, z + dz]$
in the $i$th tomography bin is
defined as
\begin{equation}
n_i(z) = \int^{z_{\rm ph}^{i+1}}_{z_{\rm ph}^i} d z_{\rm ph} \ n_{\rm gal}(z) \
p(z_{\rm ph} | z)\, ,
\end{equation}
where the function $p(z_{\rm ph}|z)$ is the probability
that a given $z$ will be measured as $z_{\rm ph}$.
This formulation assumes that $n_{\rm gal}(z)$ will be accurately
determined by an independent, dedicated
spectroscopic survey (an alternative formulation for the case of a
photometrically derived $n_{\rm gal}(z)$ can be found in \cite{Huterer:2005ez}).
For simplicity,
we choose $p(z_{\rm ph}|z)$ to be a Gaussian,
\begin{equation}
  p(z_{\rm ph}|z) = \frac{1}{\sqrt{2 \pi} \sigma_z} \exp \left[-\frac{(z_{\rm ph}+z_{\rm bias}-z)^2}
  {2 \sigma_z^2} \right]\, ,
\end{equation}
where $\sigma_z(z)$ and $z_{\rm bias}(z)$ are the scatter and the bias parameters respectively.
Following \cite{Ma:2005rc}, we discretise $\sigma_z(z)$ and $z_{\rm bias}(z)$ into
$N_{\rm pz}=31$ sets of parameters $\{\sigma_z^i,z_{\rm bias}^i\}$, where
$\sigma_z^i \equiv \sigma_z(z_i)$ and $z_{\rm bias}^i \equiv z_{\rm bias}(z_i)$,
on an equally spaced grid from $z=0$ to $z=3$.  Later on, each of $\sigma_z^i$ and $z_{\rm bias}^i$
will be treated as a free parameter in our error forecast analysis.

The exact forms of $\sigma_z(z)$ and $z_{\rm bias}(z)$ depend
on how the photometric redshift is measured by the lensing survey.
For example, $\sigma_z$ can be reduced by going to space
(to eliminate the problem of absorption in the Earth's atmosphere),
or by implementing IR-filters (sensitive to galaxies at $1 \leq z
\leq 2$).  Another possibility is to calibrate the survey
internally by doing spectroscopy on a fraction of the source
galaxies~\cite{Ishak:2004jj}.

In this study, we choose as our fiducial model $\sigma_z^i=0.05$
and $z_{\rm bias}^i=0$, and for these values, we have verified
that the accuracy requirements for the scatter and the bias are
roughly consistent with those found in \cite{Ma:2005rc}.

\subsection{Shear measurement errors}

Technically, shear measurement is performed not on the actual galaxy
images, but on images convolved with a point spread function (PSF)
which describes the various image distortions originating from the telescope,
the camera, and, in the case of ground-based surveys, the atmosphere.
Errors in the lensing signal can therefore arise from uncorrected
spatial and temporal variations in the PSF, as well as from
the deconvolution and shear estimation method itself.
Shear measurement errors can generally be discussed under the umbrellas of
multiplicative and additive errors.

\subsubsection{Multiplicative errors}

A number of algorithms have been developed so far to estimate the shear
in an observed image, all of which must be calibrated against simulations
(see, e.g., \cite{Heymans:2005rv,Hirata:2003cv} for a summary).  Uncertainties in
the calibration can lead to the measured shear being
over- or under-estimated by a constant multiplicative factor, and hence
impact directly on the $\Omega_m$ and $\sigma_8$ parameter determination.
Shear calibration errors are estimated to be $2\%$ currently, and are
expected to be controlled to $1\%$ by the time SNAP and/or LSST start~\cite{Haakon}.

To incorporate this effect into our analysis, we follow the approach of
\cite{Huterer:2005ez}, and define the theoretical convergence power spectrum as
\begin{equation}
\hat{C}^{i j}_\ell = C^{i j}_\ell\, \times (1 + f_i + f_j) ,
\end{equation}
where $f_i$ is the irreducible part of the shear calibration error in the $i$th tomography
bin, after averaging over all directions and redshifts within the bin.  Thus, for tomography
with $n_{\rm t}$ bins, there are $n_{\rm t}$ shear bias parameters.

\subsubsection{Additive errors}

Additive errors arise when artificial shear signals are induced by,
e.g., anisotropies in the PSF.  In general, the PSF varies both spatially
and temporally, and the anisotropies incurred by these variations
must be calibrated, usually using stars in the observed fields,
for each individual image.  Current simulations suggest that additive PSF
errors can be controlled to under $10^{-3}$ \cite{Haakon}.

The effects of additive errors are more difficult to parameterise, since
the errors will generally depend on $\ell$.  A first attempt in \cite{Huterer:2005ez}
defines the total convergence power spectrum as
$\hat{C}^{i j}_\ell = C^{i j}_\ell + C^{\rm add}_\ell$, where the additive error
spectrum $C^{\rm add}_\ell$ is assumed to be a power law in $\ell$.  Within this framework,
the authors of \cite{Huterer:2005ez} conclude that additive errors must be
controlled to $\sim 10^{-5}$.
We shall not be considering additive errors in this
work, but simply note here that the size of current simulations are too
small to determine if the $10^{-5}$ accuracy requirement is realistic \cite{Haakon}.

\begin{figure}[t]
\hspace*{25mm}
\epsfxsize=10cm
\epsfbox{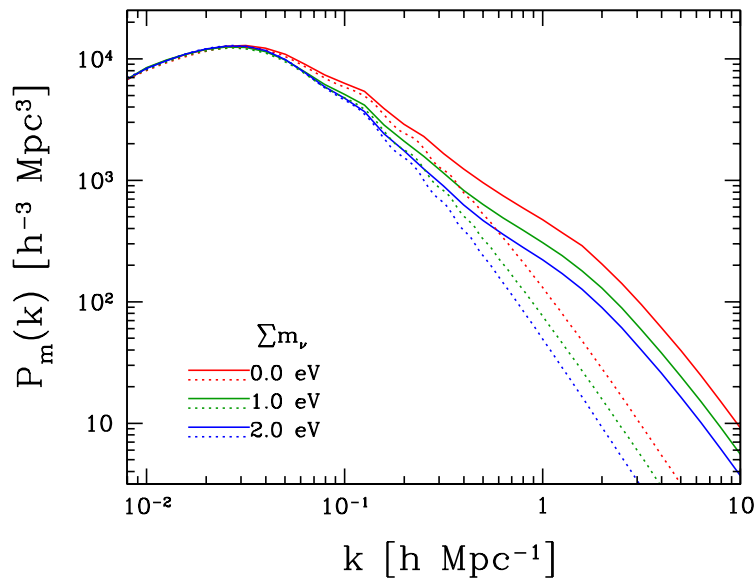}
\caption{\label{fig:nl} Nonlinear matter power spectra constructed from the halo model for various
neutrino masses (solid lines).  Also shown are the corresponding power spectra from the
linear theory (dotted lines).}
\end{figure}

\subsection{Matter power spectrum \label{sec:pnl}}

The sensitivity of weak lensing surveys peaks at scales of several arcminutes,
i.e., $\ell \sim 1000 $ \cite{Jain:1996st} (cf.\ Figure \ref{fig:cl}).
At these scales the shear statistics
is dominated by density perturbations of comoving wavenumbers
$k \sim 1 \to 10 \ h {\rm Mpc}^{-1}$
at $z \sim 0.5$.
Thus, weak lensing probes a convolution of linear and nonlinear scales at
present and earlier times, and the linear power spectra $P^{\rm L}(k)$
computed by, e.g., CMBFAST \cite{bib:cmbfast} alone will not suffice to determine
total matter power spectrum $P_m(k)$.

Assuming that baryons trace CDM exactly, accurate predictions of $P_m(k)$ can be obtained by means
of collisionless $N$-body simulations.  At present various cosmological simulation codes
agree at the $5 \to 10 \ \%$ level
\cite{Heitmann:2004gz} for $\Lambda$CDM cosmologies.
In the future, however, $P_m(k)$ will have to be calibrated to about $1 \to 2 \ \%$
accuracy in order not to compromise the cosmological parameter extraction from
ambitious surveys such as SNAP and LSST \cite{Huterer:2004tr}.  Inclusion of baryon physics introduces
an extra layer of complexity;
dissipative processes such as
radiative cooling and star formation enters at $k \gwig 1 \ h {\rm Mpc}$, and require in
addition the use of hydrodynamical simulations to accurately model the behaviour of the
baryon gas.  We note in passing that large volume simulations of dark energy cosmologies
to date
have focussed mainly on $\Lambda$CDM, although there have been some efforts in the direction of
$w \neq -1$ and/or a
time-varying equation of state
(e.g., \cite{Klypin:2003ug,Linder:2003dr,Dolag:2003ui,Bartelmann:2004gv}).  To our knowledge,
simulations of dark energy cosmologies with mixed dark matter have not yet been
reported in the literature.

A number of semi-analytical methods and fitting formulae have been
devised ~\cite{Peacock:1996ci,Seljak:2000gq,Peacock:2000qk,Ma:2000ik,Smith:2002dz},
some of which tested to roughly $10 \ \%$ accuracy against $N$-body
simulations \cite{Ishak:2003zw,White:2003xz},
to map between the linear and nonlinear power spectra
in CDM
cosmologies.
Virtually all weak lensing parameter constraint analyses to date
employ one of these mapping methods. In this paper, we correct for
nonlinearity using the halo model
\cite{Seljak:2000gq,Peacock:2000qk,Ma:2000ik}, assuming that it
can be generalised to general dark energy and mixed  dark matter cosmologies.
Here, the total matter power spectrum is taken to be a weighted average of the
neutrino ($\nu$), and the combined baryon ($b$) and CDM ($c$) power spectra,
\begin{equation}
\label{eq:pm}
P_m(k) = \left[f_{\nu} \sqrt{P^{\rm L}_{\nu}(k)}+(f_b+f_c) \sqrt{P_{b+c}^{\rm NL}(k)}\right]^{2},
\end{equation}
with weights $f_i=\Omega_i/\Omega_m$, $i=\nu,b,c$,
and  $\Omega_m \equiv \Omega_{\nu}+\Omega_b+\Omega_c$.
The nonlinear baryon and CDM power spectrum $P_{b+c}^{\rm NL}(k)$ is mapped from its linear
counterpart,
\begin{equation}
\label{eq:pbc}
P^{\rm L}_{b+c}(k) = (f_b+f_c)^{-1} \left[f_b \sqrt{P^{\rm L}_b(k)}+f_c \sqrt{P^{\rm L}_c(k)} \right]^{2},
\end{equation}
following the halo model prescription.  See \ref{sec:halomodel} for details.

Figure \ref{fig:nl} shows
the nonlinear matter power spectra constructed in this fashion for various neutrino masses.
Clearly, the suppression of power due to neutrino free-streaming extends well into the
nonlinear regime ($k \gwig 0.2 \ h {\rm Mpc}$).  Indeed, the halo model predicts an even stronger suppression
on nonlinear scales than in the linear region ($k \lwig 0.2 \ h {\rm Mpc}$).
However, compared with their corresponding uncorrected power spectra (i.e., power spectra
calculated from the  linear theory, extended into the nonlinear regime),
this relative suppression, i.e., the ratio $\Delta P_m(k)/P_m(k)$, in the corrected spectra
seems to be somewhat weaker.  (Observe in Figure \ref{fig:nl} that the gaps between the solid lines seem to
be narrower than those between the dotted lines at a given $k$.)
This result remains to be verified by simulations.

 We do not consider nonlinearity
in the neutrino power spectrum  $P^{\rm L}_{\nu}(k)$ incurred from late-time infall into existing halos
\cite{Ringwald:2004np}, since it affects the total matter power spectrum and hence the weak lensing
observables only at the $\sim 0.1 \%$ level \cite{Abazajian:2004zh,Hannestad:2005bt}.
We ignore also corrections due to cooling baryons and the intra-cluster gas
\cite{White:2004kv,Zhan:2004wq}, although results from recent $N$-body/hydrodynamical simulations
suggest that baryon physics affects the weak lensing signal at the
$\lwig 0.5 \ \%$ and the $1 \to 10 \ \%$ levels at $\ell < 1000$ and $1000 < \ell < 10000$
respectively \cite{Jing:2005gm}.

\section{Error forecast \label{sec:forecast}}

We use the Fisher matrix formalism to estimate the constraints on
various cosmological parameters from weak lensing tomography.

\subsection{Fisher matrix}

The Fisher matrix is defined as
\begin{equation}
F_{\alpha \beta} = - \left\langle \frac{\partial^2 \ln L}{\partial p_{\alpha} \partial p_{\beta}}
\right\rangle,
\end{equation}
where $L({\bm x},{\bm p})$ is the likelihood function of the data set ${\bm x}$ given the
parameters ${\bm p}=\{p_1,\ldots,p_{\alpha}\}$ of the theoretical model.  When evaluated in the vicinity of
the fiducial model, the Fisher matrix quantifies the best statistical error on the parameter $p_{\alpha}$
achievable with a given data set via
\begin{equation}
\sigma(p_{\alpha}) \geq \sqrt{({\bm F}^{-1})_{\alpha \alpha}},
\end{equation}
where $\sigma(p_{\alpha})$ denotes the $1 \sigma$ error on $p_{\alpha}$ after
marginalisation over all other parameters.

\begin{table}
\caption{\label{table:wl} Generic weak lensing survey parameters.}
\begin{indented}
\item[]\begin{tabular}{@{}lcccc}
\br
 & $f_{\rm sky}$ & $z_0$ & $\gamma_{\rm rms}$ & $\bar{n}_{\rm gal}$ (${\rm arcmin}^{-2}$)  \\
\mr
Wide, LSST-like & 0.7 & 1.0 & 0.4 & 30 \\
Deep, SNAP-like & 0.01 & 1.5 &  0.25 & 100 \\
\br
\end{tabular}
\end{indented}
\end{table}

Neglecting non-Gaussian corrections, the Fisher matrix for weak lensing tomo-graphy has the form
\begin{equation}
\label{eq:fisher}
F_{\alpha \beta} = \sum^{\ell_{\rm max}}_{\ell_{\rm min}} (\ell + 1/2) f_{\rm sky} \ \Delta \ell \
{\rm Tr} \left( \widetilde{\bm C}_{\ell}^{-1} \frac{\partial {\bm C}_{\ell}}{\partial p_{\alpha}}
\widetilde{\bm C}_{\ell}^{-1} \frac{\partial {\bm C}_{\ell}}{\partial p_{\beta}} \right).
\end{equation}
Here, we take the lensing observable ${\bm C}_{\ell}$ to be the number weighted convergence
power spectrum ${\bm C}_{\ell} \doteq \bar{n}_i \bar{n}_j \hat{C}^{ij}_{\ell}$, since
it contains more information than does the binned spectrum
$\hat{C}^{ij}_{\ell}$ to reflect changes in the photometric redshift parameters
\cite{Ma:2005rc}.
The variance  $\widetilde{\bm C}_{\ell}$
is correspondingly defined as a number weighted sum of the convergence and the noise power
spectra, $\widetilde{\bm C}_{\ell} \doteq  \bar{n}_i \bar{n}_j (\hat{C}^{ij}_{\ell} +
\delta_{ij} \gamma_{\rm rms}^2/\bar{n}_{i}) $.  Note that
in the case of fixed photometric redshift parameters, these definitions lead to the same $F_{\alpha \beta}$
as their more common representations,
${\bm C}_{\ell} \doteq \hat{C}^{ij}_{\ell}$ and
$\widetilde{\bm C}_{\ell} \doteq  \hat{C}^{ij}_{\ell} +
\delta_{ij} \gamma_{\rm rms}^2/\bar{n}_{i}$.  We consider a variety of tomography bin numbers
($n_{\rm t}=1,3,5,8$),  each with equal redshift binning $\Delta z$ between $z=0$ and  $z=3$
(e.g., $\Delta z=0.6$ for $n_{\rm t}=5$).

The minimum and maximum multipoles $\ell_{\rm min}$ and
$\ell_{\rm max}$ are taken to be $50$ and $3000$ respectively,
the latter so as to avoid the highly non-Gaussian signals at the very
small scales.
Between $\ell_{\rm min}$
and $\ell_{\rm max}$ we use $100$ logarithmic bins of width $\Delta \ell$; variations in the
 $\ell$  binning have
little effect on our results, since the convergence power
spectrum is fairly featureless.
The other parameters---the sky coverage $f_{\rm sky}$,
the surface density of galaxies
$\bar{n}_{\rm gal} = \sum_i \bar{n}_i$, and their rms
intrinsic ellipticity $\gamma_{\rm rms}$---are generally
survey-dependent.   In our analysis, we consider two generic weak lensing
surveys: a ground-based LSST-like ``wide'' survey, and a space-based SNAP-like
``deep'' survey,
with  survey parameters displayed in Table \ref{table:wl}.

We are primarily interested in how parameter constraints derived
from CMB data can be improved with the inclusion of weak lensing.
Since the primary CMB anisotropies and the galaxy shears probed by
weak lensing surveys are generated at two vastly different epochs, we may take them
to be uncorrelated.  Thus the total Fisher matrix is simply the sum of two independent
matrices,
\begin{equation}
\label{eq:totalfisher}
F_{\alpha \beta}^{\rm total} = F_{\alpha \beta}^{\rm WL} + F_{\alpha \beta}^{\rm CMB},
\end{equation}
which is a good approximation provided we confine the analysis to the
{\it unlensed} CMB anisotropies.  (Weak lensing of the CMB and of galaxies are necessarily
strongly
correlated since the lensing signals are generated by much of the same
intervening large scale structure.)

\begin{table}
\caption{\label{table:planck} Experimental specifications for Planck, assuming one year of observation.}
\begin{indented}
\item[]\begin{tabular}{@{}ccccc}
\br
$f_{\rm sky}$ & $\nu$ (GHz)  & $\theta_b$ (arcmin) & $\Delta_T$ ($10^{-6} \ T$) & $\Delta_E$ ($ 10^{-6}\ T$) \\
\mr
0.65 &100 & 9.5 & 2.5 & 4.0 \\
& 143 & 7.1 & 2.2 & 4.2 \\
& 217 & 5.0 & 4.8 & 9.8 \\
\br
\end{tabular}
\end{indented}
\end{table}

The Fisher matrix for the unlensed CMB anisotropies $F_{\alpha \beta}^{\rm CMB}$
is given likewise by equation (\ref{eq:fisher}), but with $\widetilde{\bm C}_{\ell} =
{\bm C}_{\ell} + {\bm N}_{\ell}$, and
\begin{equation}
{\bm C}_{\ell} = \left( \begin{array}{cc}
                        C^{TT}_{\ell} & C^{TE}_{\ell} \\
                        C^{TE}_{\ell} & C^{EE}_{\ell}
                        \end{array} \right), \quad
{\bm N}_{\ell} = \left( \begin{array}{cc}
                        N^{TT}_{\ell} & 0 \\
                        0 & N^{EE}_{\ell}
                        \end{array} \right),
\end{equation}
where $C^{TT}_{\ell}$, $C^{EE}_{\ell}$, and $C^{TE}_{\ell}$ are the
auto and cross power spectra of the CMB temperature and $E$-type polarisation.
These are computed using  CMBFAST version 4.5.1 \cite{bib:cmbfast}.
We do not consider $B$-type polarisation, although it will most likely be measured by
the next generation of CMB probes.  The noise power spectra ${\bm N}_{\ell}$
vary with the detector performance according to
\begin{equation}
N^{aa}_{\ell} =  \left[ \sum_{\nu} (N_{\ell,\nu}^{aa})^{-1} \right]^{-1}, \quad
N_{\ell,\nu}^{aa} = (\theta_b \Delta_a)^2 \exp[\ell (\ell +1) \theta_b^2/8 \ln 2],
\end{equation}
where $\nu$ denotes the detector frequency channel, $\theta_b$ is the beam width, and $\Delta_a$ is
the temperature/polarisation sensitivity per pixel.  Table \ref{table:planck} shows the expected
performance of the Planck satellite \cite{bib:planck} (assuming one year of observation),
and we take $\ell_{\rm max}=1500$ in our analysis.

Prior information on the photometric redshift uncertainties (i.e., the parameters $z_{\rm bias}^i$ and $\sigma_z^i$)
and the shear calibration bias (i.e., $f_i$) can be included in the analysis by
adding  the diagonal matrix
$F_{\alpha \beta}^{\rm prior} = \delta_{\alpha \beta} \ \sigma(p_{\alpha})^{-2}$ to
the total Fisher matrix (\ref{eq:totalfisher}).

\begin{figure}[t]
\epsfxsize=15cm \epsfbox{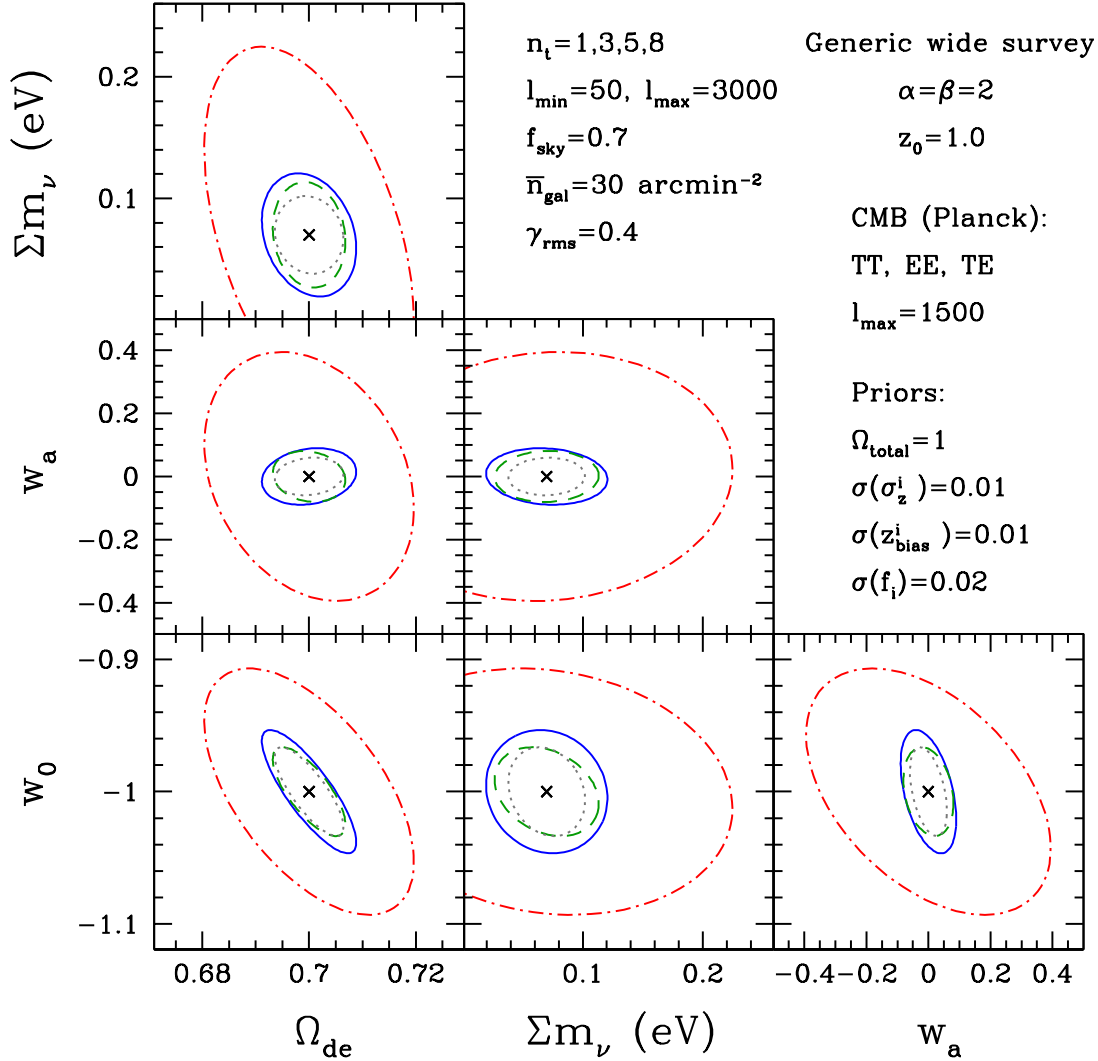}
\caption{\label{fig:wide}  Projected $1\sigma$ contours in the
dark energy and neutrino parameter space from
an LSST-like ground-based wide lensing survey (with and without tomography) and Planck
for an eleven parameter cosmological model.  The red (dot-dash), blue (solid),
green (dashed), and gray (dotted) lines correspond to one, three, five, and eight
tomography bins respectively.
Each contour
comes from marginalising over the
other nine cosmological parameters and
all $2\times N_{\rm pz}+n_{\rm t}$ systematic parameters, with Gaussian priors
imposed on the latter.}
\end{figure}

\begin{figure}[t]
\epsfxsize=15cm \epsfbox{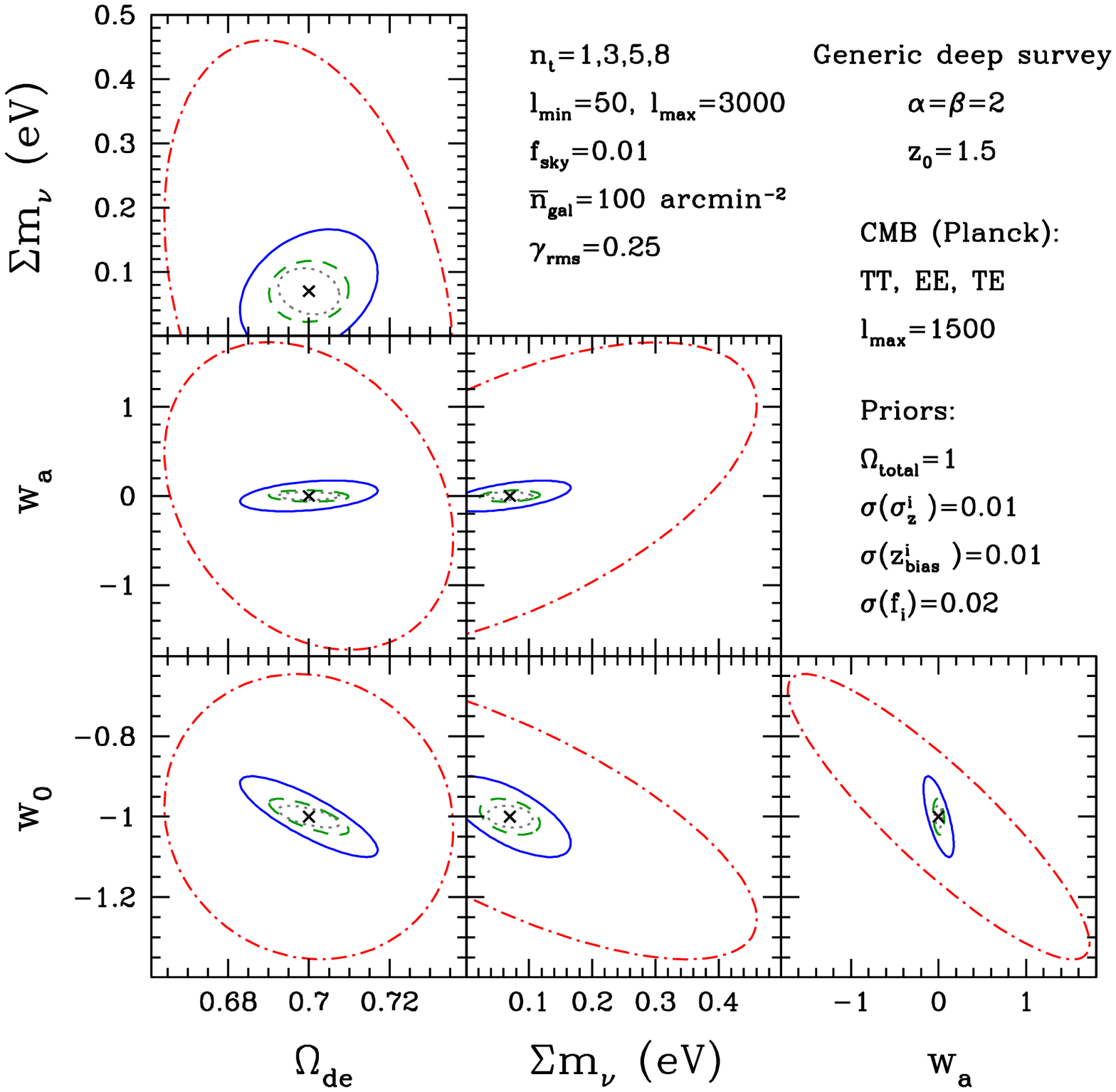}
\caption{\label{fig:deep} Same as Figure \ref{fig:wide}, but from
a SNAP-like space-based deep survey and Planck.}
\end{figure}

\subsection{The fiducial model\label{sec:fid}}

We consider a spatially flat cosmological model with eleven parameters,
the  fiducial values of which are as follows:
$\sum m_{\nu}=0.07 \ {\rm eV}$, $\Omega_{\rm de} = 0.7$, $\Omega_b h^2 = 0.0245$,
$\Omega_c h^2 = 0.1225$, $w_0 = -1$, $w_a=0$, $\tau=0.05$, $n_s=1$, $\alpha_s=0$, $\sigma_8 = 0.9$, and
$N_{\rm eff}=3$.  In this model, the time-dependent dark energy equation of state is parameterised
as \cite{Chevallier:2000qy,Linder:2002et}
\begin{equation}
w(z)=w_0+w_a [1-a(z)].
\end{equation}
See, e.g., \cite{Hannestad:2004cb,Wang:2004py,Upadhye:2004hh} for alternative
parameterisations.
We also allow for nonstandard relativistic degrees of
freedom $N_{\rm eff} \neq 3$, and a running scalar spectral index $\alpha_s$
evaluated at $k_0=0.05 \ h {\rm Mpc}^{-1}$.  These extra parameters,
$\{N_{\rm eff},\alpha_s,w_0,w_a\}$, are known to be
degenerate with the neutrino mass $m_{\nu}$.  For simplicity, we take the number of massive neutrino
species to be one.
Spatial flatness implies $\Omega_{\rm de} +\Omega_{\nu}+\Omega_b+\Omega_c=1$.

In addition to the eleven cosmological parameters, we marginalise over $2\times N_{\rm pz}+n_{\rm t}$
nuisance parameters, $\sigma_z^i$, $z_{\rm bias}^i$, and $f_i$.  The fiducial values of these
parameters are taken to be $\sigma_z^i=0.05$, $z_{\rm bias}^i=0$, and $f_i=0$, and the
Gaussian priors $\sigma(\sigma_z^i) = 0.01$, $\sigma(z_{\rm bias}^i)=0.01$, and $\sigma(f_i)=0.02$
are used in the marginalisation.
We do not employ priors on  the cosmological parameters.

\begin{table}
\caption{\label{table:constraints} Projected $1 \sigma$ errors from various combinations
of Planck and weak lensing.  The labels ``Wide'' and ``Deep'' denote the two generic lensing
surveys considered in this work, and the suffix number indicates the number of tomography bins
used in the analysis.}
\begin{indented}
\item[]\begin{tabular}{@{}lccccc}
\br
& Planck  & +Wide-1 & +Wide-5 & +Deep-1 & +Deep-5 \\
\mr
$\sigma(\sum m_{\nu})$ (eV) & 0.48 & 0.15 & 0.043 & 0.39 & 0.047 \\
$\sigma(\Omega_{\rm de})$ & 0.08 & 0.020 & 0.0068 & 0.036 & 0.0099 \\
$\sigma(\Omega_b h^2)$ & 0.00028 & 0.00016 & 0.00013 & 0.00024 & 0.00014 \\
$\sigma(\Omega_c h^2)$ & 0.0026 & 0.0017 & 0.0015 & 0.0019 & 0.0015\\
$\sigma(w_0)$ & 0.83 & 0.093 & 0.034 & 0.35 & 0.045 \\
$\sigma(w_a)$ & 4.0 & 0.39 & 0.081 & 1.7 & 0.063 \\
$\sigma(\tau)$ & 0.0046 & 0.0043 & 0.0042 & 0.0045 & 0.0043\\
$\sigma(n_s)$ & 0.0089 & 0.0056 & 0.0028 & 0.0074 & 0.0047\\
$\sigma(\alpha_s)$ & 0.024 & 0.013 & 0.0061 & 0.020 & 0.012 \\
$\sigma(\sigma_8)$ & 0.084 & 0.019 & 0.0076 & 0.030 & 0.0092 \\
$\sigma(N_{\rm eff})$ & 0.19 & 0.11 & 0.067 & 0.14 & 0.093\\
\br
\end{tabular}
\end{indented}
\end{table}

\subsection{Results}

Figures \ref{fig:wide} and \ref{fig:deep} show the projected $1\sigma$ contours in the
dark energy and neutrino parameter space expected from
two generic weak lensing surveys, with and without tomography, in combination with
prospective CMB data from the Planck mission. (See Table \ref{table:constraints}
for all projected errors, and Table \ref{tab:correlation} for the
parameter error correlation matrix
$F_{\alpha \beta}/\sqrt{F_{\alpha \alpha} F_{\beta \beta}}$.)
Clearly, the addition of tomographic binning significantly reduces the errors for all four parameters,
$\{\sum m_{\nu},w_0,w_a,\Omega_{\rm de}\}$.  Focussing on the LSST-like
wide survey (Figure~\ref{fig:wide}), we see that the improvement on the $w_a$ error
estimate is roughly a factor of five.
For the neutrino mass, the gain is slightly more moderate: roughly a
factor of three difference between the cases of no tomography and tomography with
five bins.  This follows from the fact that
tomography does not directly add extra information on the shape of the matter power
spectrum (see discussion in section \ref{sec:denu}).

\begin{table}
\caption{\label{tab:correlation}  Parameter error correlation matrices for
(a) Planck+Wide-1,-5, and (b) Planck+Deep-1,-5.  In each case,
the top triangular matrix corresponds to tomography with five bins, while
the bottom triangle pertains to lensing without tomography.  Photometric redshift
errors have been omitted for the lack of space.}
\vspace*{2mm}
\hspace{25mm} \includegraphics[angle=90,height=213mm]{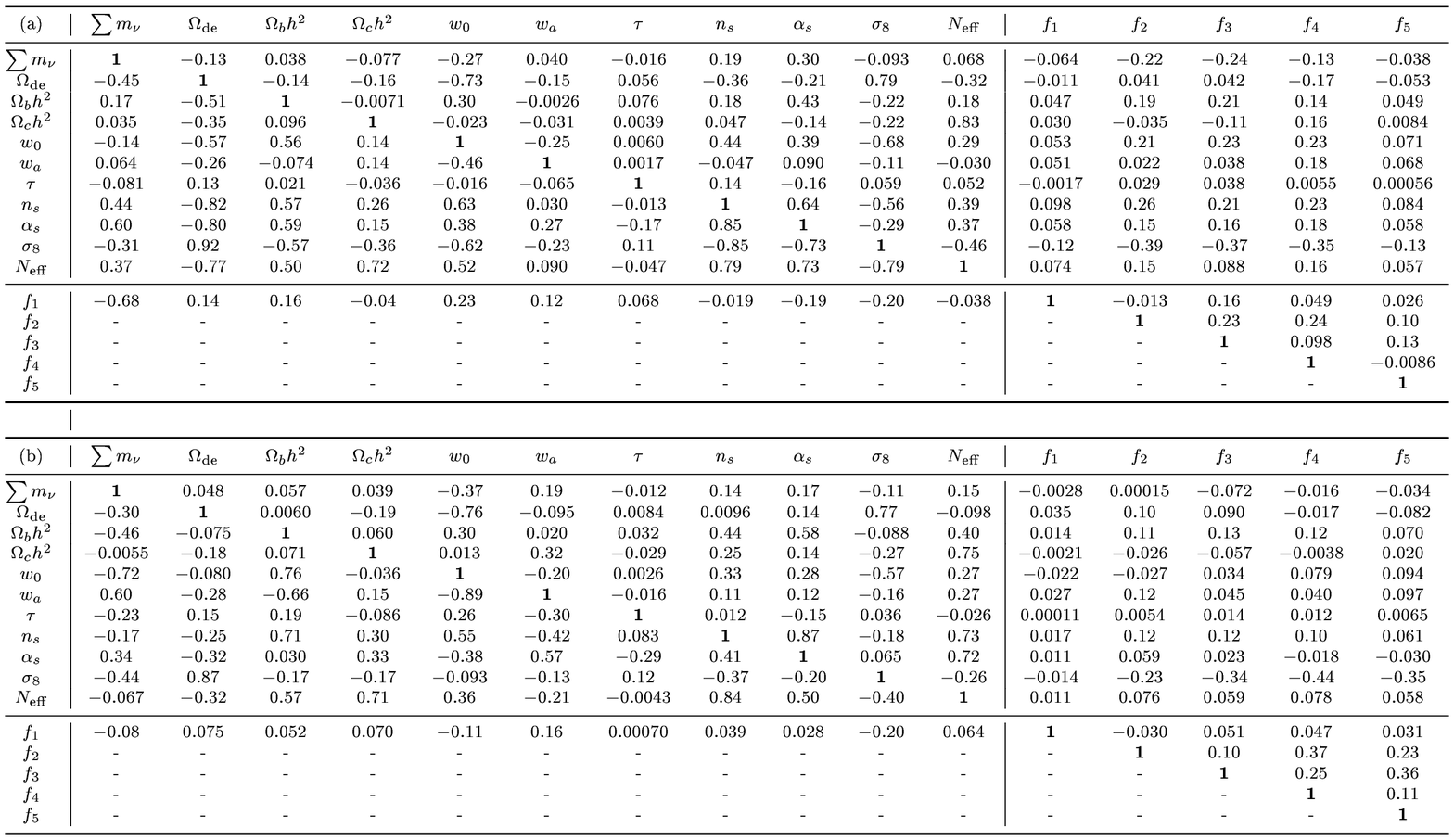}
\end{table}

The SNAP-like deep survey (Figure \ref{fig:deep}) offers even more
significant relative improvement,  since it probes further back in redshift where
the effect of dark energy on the distance--redshift relation is more prominent
(cf.\ Figure~\ref{fig:weights}).  However, owing to its limited sky coverage,
the absolute errors (with tomography) are comparable to those from the wide survey.
Should the value of $f_{\rm sky}$ be increased, one can scale the errors presented
here roughly via
\begin{equation}
\sigma(p_{\alpha},f_{\rm sky}) \sim (f_{\rm sky,fid}/f_{\rm sky})^{1/2}
\sigma(p_{\alpha},f_{\rm sky,fid}),
\end{equation}
where $f_{\rm sky,fid}$ denotes the sky coverage of the
fiducial survey considered in this work.  This expression is applicable for
$\{\sum m_{\nu},w_0,w_a,\Omega_{\rm de}\}$ with tomography,
since the constraints on these parameters come predominantly from lensing, not from
CMB.

\begin{figure}[t]
\epsfxsize=15.5cm \epsfbox{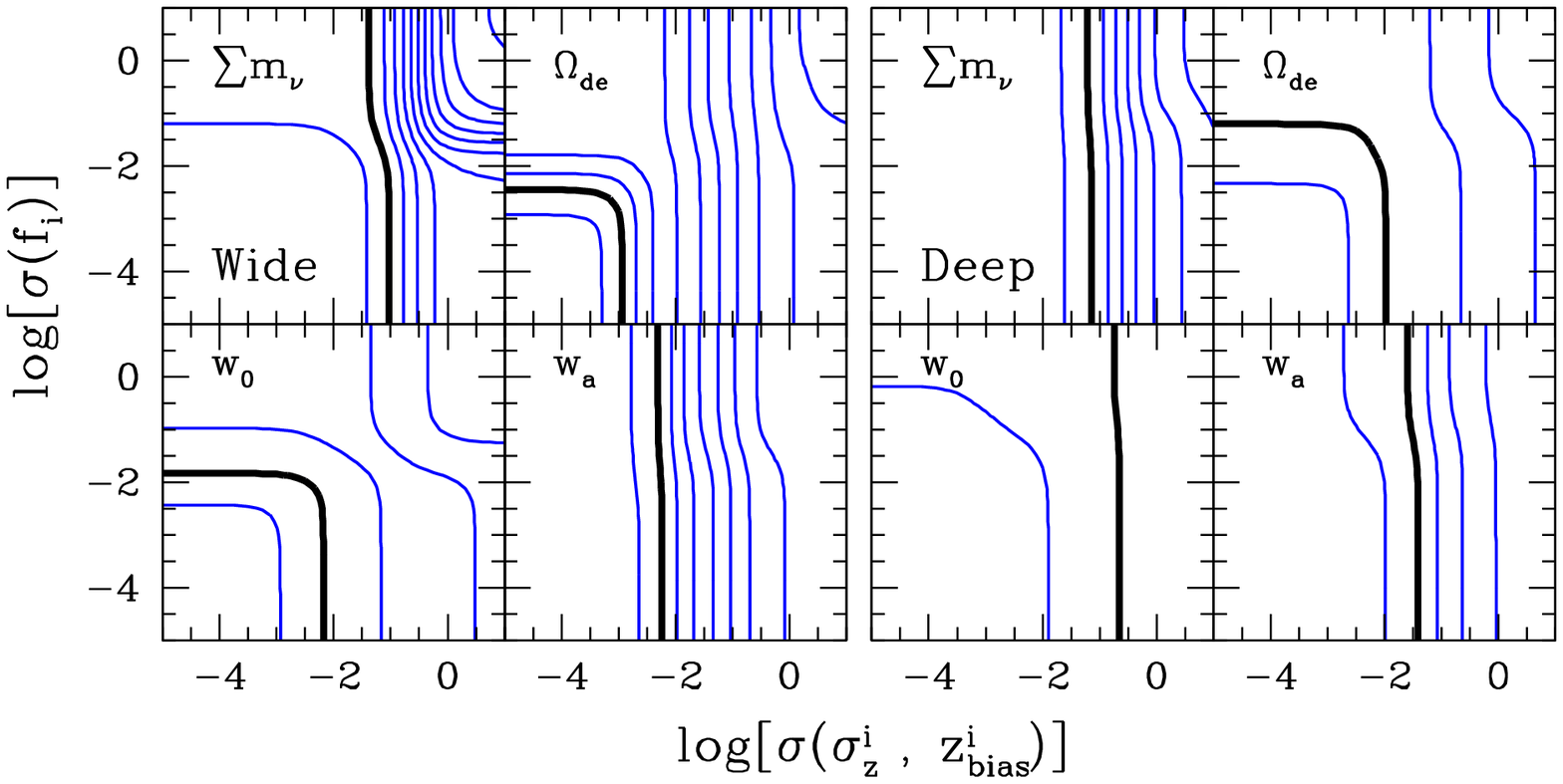}
\caption{\label{fig:degradation} Degradation in the projected sensitivities,
relative to the best achievable errors by a systematics-free lensing survey,
as a function of priors on the shear calibration bias
and photometric redshift errors.  Contours of constant degradation are shown in
intervals of $20 \ \%$, in ascending order from left/bottom to right/top.
The thick black line in each plot indicates a sensitivity
degradation of $30 \ \%$.  {\it Left}: Contours for Planck+Wide-5.
{\it Right}: Contours for Planck+Deep-5.}
\end{figure}

It is interesting to compare these constraints with those derived from CMB alone.
Table \ref{table:constraints} shows the projected errors on various cosmological parameters
from Planck alone, and in combination with a selection of
lensing surveys. CMB alone has very little sensitivity to the dark energy equation
of state; lensing is crucial for the determination of $w_0$ and $w_a$.
On the other hand, although the projected error on the neutrino mass from Planck alone is
comparable to what can be achieved today with CMB+LSS+SNIa+BAO, weak lensing
(without tomography) brings it down to a level
comparable to the projected sensitivities of future galaxy surveys
(e.g., \cite{Lesgourgues:2004ps}).  The addition of tomography binning pushes the
error down even further.

Of course, these impressive sensitivities come at a price: appropriate
priors have been imposed on the systematics parameters in order not to degrade the
sensitivity to any one (cosmological) parameter by more than $\sim 50 \ \%$ relative to
the best achievable error by a systematics-free lensing survey.
Figure \ref{fig:degradation} shows the degradation in the projected sensitivities
to $\{\sum m_{\nu},w_0,w_a,\Omega_{\rm de}\}$ as a function of priors on the shear
calibration bias and photometric redshift errors for two combinations of Planck
and lensing tomography with five bins.  Clearly,
the largely vertical contours suggest that the accuracy to which
the photometric redshift uncertainties can be controlled
is a crucial factor in deciding a survey's parameter sensitivities.
For the neutrino mass determination, for example, the sensitivity degrades
rapidly as the accuracy requirement for $\{\sigma_z^i,z_{\rm bias}^i\}$ is relaxed.
In contrast, the degradation due to shear calibration errors saturates at $\sim 70 \ \%$
for $\Omega_{\rm de}$ (Planck+Wide-5), and much less for other parameters.  This
is qualitatively consistent with the findings of \cite{Huterer:2005ez}, where the
shear calibration errors are said to enter a ``self-calibration'' regime.

Changing $\ell_{\rm max}$ to a lower value degrades further the parameter
sensitivities.  For the neutrino mass extraction, the degradation is
$\sim 30 \ \%$ from $\ell_{\rm max}=3000$ to $\ell_{\rm max}=1000$.
For the dark energy parameters, on the other hand, the sensitivities 
degrade generally by less than $20 \ \%$.

Focussing now on the $m_\nu,w$ degeneracy,
Table \ref{table:blah} compares various projected constraints on the neutrino mass
for the eleven parameter model (cf.\ section \ref{sec:fid}) and
a seven parameter model which assumes $\{N_{\rm eff},\alpha_s,w_0,w_a\}$ to be fixed at
$\{3,0,-1,0\}$.  From Planck and weak lensing without tomography,
the expected $1 \sigma$ error for the two models, $0.15 \ {\rm eV}$ and $0.082 \ {\rm eV}$,
differ by almost a factor of two.  With the inclusion of five tomography  bins,
the errors reduce to $0.043 \ {\rm eV}$ and $0.037 \ {\rm eV}$ respectively,
and only a $\sim 15 \ \%$ discrepancy remains between them.  This clearly
demonstrates that weak lensing tomography can break
the $m_{\nu},w$ degeneracy very effectively.

\section{Conclusions \label{sec:conclusions}}

Cosmological probes are already at an accuracy level to probe
important aspects of particle physics such as neutrino masses
and the dark energy equation of state.  Current CMB, LSS, SNIa
and BAO data can already pin the former down to
$\sum m_{\nu} \lwig 0.5 \ {\rm eV}$~\cite{Goobar:2006xz}.
With future CMB data and weak lensing surveys, the projected
sensitivity to these parameters
will be far better.  From a combined analysis of prospective
data from the Planck mission and a future LSST-like
full-sky lensing survey, the sensitivity to $\sum m_{\nu}$ is expected to
reach $0.05 \ {\rm eV}$,
 sufficient to probe
the difference between the normal and inverted hierarchy schemes.

\begin{table}
\caption{Projected $1\sigma$ constraints on the neutrino mass from
various combinations of Planck and an LSST-like wide lensing survey
(with one and five tomography bins) for two cosmological models.
The eleven parameter model corresponds to that presented in section
\ref{sec:fid}, while the seven parameter model assumes
$\{N_{\rm eff},\alpha_s,w_0,w_a\}$ to be fixed at
$\{3,0,-1,0\}$.\label{table:blah}}
\begin{indented}
\item[]\begin{tabular}{@{}llc}
\br
Model & Cosmological probes & $\sigma(\sum m_{\nu})$\\
\mr
11 parameters & Planck only & $0.48 \ {\rm eV}$ \\
11 parameters & Planck+Wide-1 & $0.15 \ {\rm eV}$ \\
11 parameters & Planck+Wide-5 & $0.043 \ {\rm eV}$ \\
7 parameters & Planck+Wide-1 & $0.082 \ {\rm eV}$ \\
7 parameters & Planck+Wide-5 & $0.037 \ {\rm eV}$ \\
\br
\end{tabular}
\end{indented}
\end{table}

It is interesting to compare our results to those derived in
\cite{Song:2003gg}, which also considers the combination of
Planck data with a future weak lensing survey.
For the model closest to our Planck+Wide-5 eleven parameter scenario
(cf.\ Table \ref{table:blah}), they find a sensitivity of $\sigma(\sum m_\nu) =
0.045 \ {\rm eV}$, compared to our $0.043 \ {\rm eV}$.  Systematics issues
in the weak lensing survey have been neglected in the analysis of
\cite{Song:2003gg}.  The same is true in our analysis
in the sense that we have assumed the systematic effects
to be controlled to the accuracy  needed in order not to
seriously degrade the parameter determination.

A similar sensitivity to $\sum m_\nu$ can also be reached by combining Planck data
with future cluster abundance measurements  \cite{Wang:2005vr},  $\sigma(\sum m_\nu) =
0.034 \ {\rm eV}$,
and high-redshift galaxy surveys \cite{Takada:2005si}, $\sigma(\sum m_\nu) =
0.059 \ {\rm eV}$ ($0.5 < z < 2$) and  $0.043 \ {\rm eV}$ ($2< z <4$).  Note that these numbers were derived in \cite{Wang:2005vr,Takada:2005si}
for an equivalent of our seven parameter model.  However,
since both methods provide a measure of the distance--redshift
relation, and therefore share the same principle with weak lensing
 tomography, it is very probable that these alternative probes could also offer
some improvement towards the degeneracy between the neutrino mass and the dark energy equation of state,
should the latter have been treated as a free parameter.  This remains to be verified.

A partial breakage of the said degeneracy is also possible
by cross-correlating CMB and LSS measurements.  The analysis of
\cite{Ichikawa:2005hi} finds that the combination of Planck and future
LSS data, including cross-correlation between the two data sets,
can bring the $\sum m_\nu$ sensitivity to the $0.1 \ {\rm eV}$ level.
This, however, is not competitive with what is expected from weak
lensing surveys.

Finally, the weak lensing signal can be extracted
from the CMB data itself. With data from the Planck satellite, the
authors of~\cite{Lesgourgues:2005yv} estimate that the neutrino
mass can be determined to $\sigma(\sum m_\nu) = 0.51 \ {\rm eV}$ without
lensing extraction, a number comparable to our $0.48 \ {\rm eV}$ (cf.\ Table \ref{table:blah}).
With lensing extraction the error reduces to $0.15 \ {\rm eV}$, i.e.,
a factor of three improvement.  Clearly, CMB lensing adds some of the same
information gained in a galaxy lensing survey, except that the source
redshift is locked at $z \sim 1100$. Therefore, a CMB survey with
a much higher resolution is needed in order to breach the
$\sigma(\sum m_\nu) = 0.1 \ {\rm eV}$ barrier for the neutrino mass
determination.  With the proposed Inflation Probe mission, the analysis of
\cite{Lesgourgues:2005yv} finds  that the sensitivity could reach
$0.04 \ {\rm eV}$ if lensing extraction is included, a precision similar to
that attainable by large scale tomographic galaxy lensing surveys  found in our analysis.
The combination of these techniques could plausibly lead to even
better sensitivities.

\section*{Acknowledgements}

We thank H{\aa}kon Dahle and Bjarne Thomsen for enlightening
discussions on the systematics related to weak lensing surveys.

\appendix

\section{The halo model \label{sec:halomodel}}

The halo model supposes that all matter in the universe is
partitioned
into distinct units, the halos.  This assumption allows one to study
the universal matter distribution in two steps: the distribution of matter
within each halo, and the spatial distribution of the halos themselves.
In its simplest application,
the halo model proposes that the matter power spectrum $P^{\rm NL}(k,z)$
be composed of two distinct terms
(e.g., \cite{Cooray:2002di}),
\begin{equation}
\label{eq:halomodel}
P^{\rm NL}(k,z) = P^{\rm 1-halo} (k,z) + P^{\rm 2-halo}(k,z),
\end{equation}
where
\begin{eqnarray}
\label{eq:p1h&p2h}
P^{\rm 1-halo} (k,z) &=& \int d M \ n(M,z)
\left(\frac{M}{\bar{\rho}_{\rm halo}} \right)^2
|u(k|M)|^2,\nonumber \\P^{\rm 2-halo} (k,z) &=& P^{\rm L}(k,z) \left[ \int d M
\ n(M,z)\ b(M) \left(\frac{M}{\bar{\rho}_{\rm halo}} \right) u(k|M)
\right]^2,
\end{eqnarray}
describe, respectively, the correlations between two
points drawn from the same halo and from two different halos.
The former dominates on small scales, while
the latter rises to prominence on large scales  and
approaches the power spectrum calculated from
linear theory $P^{\rm L}(k,z)$ as $k \to 0$.
The average  matter density $\bar{\rho}_{\rm halo}$ counts all
matter clustered in halos.
For a basic $\Lambda$CDM set-up, $\bar{\rho}_{\rm halo}$ is
well approximated by
$\bar{\rho}_{\rm halo} \simeq \bar{\rho}_m
\equiv
\Omega_m \rho_{\rm crit} $,
where $\rho_{\rm crit}$ is the present critical density.

Three pieces of information are required to complete the model.
\begin{enumerate}
\item The mass function $n(M,z)$, usually written as
$n(M,z) \ d M = \bar{\rho}_{\rm halo}/M \ f(\nu)  \ d \nu$,
 specifies the
 number density of halos of mass $M$ at redshift $z$.
Here, $f(\nu)$ is a universal function of the peak height
$\nu = \delta_{\rm sc}^2/\sigma_{\rm L}^2(M,z)$, where
$\delta_{\rm sc}=1.68$ is the linear overdensity at the epoch of
spherical collapse, and
\begin{equation}
\label{eq:sigma}
\sigma_{\rm L}^2 (M,z)\equiv \int \frac{d^3k}{(2 \pi)^3} P^{\rm L}(k,z) |W(k
R)|^2
\end{equation}
is the rms linear fluctuations filtered  with a
tophat function $W(x) = 3/x^3 (\sin x - x \cos x)$ on a scale
of $R=(3 M/4 \pi \bar{\rho}_m)^{1/3}$.
A number of variants for $f(\nu)$ exists in the
literature.
Here, we adopt the version proposed by Sheth and Tormen \cite{Sheth:1999mn}.

\item The linear bias $b(\nu)$ parameterises the clustering
strength of halos relative to the underlying dark matter, and obeys
$\int d \nu \ f(\nu) \ b(\nu)=1$ by construction.

\item The function $u(k|M)$ is the Fourier transform
of the matter distribution
$\rho(r|M)$ within a halo, normalised to the halo mass $M$.
A natural choice for $\rho(r|M)$ would be the
 Navarro--Frenk--White profile \cite{Navarro:1995iw,bib:nfw}.
\end{enumerate}

To generalise the halo model to mixed dark matter cosmologies, we assume
that only baryons and CDM cluster in halos, while the neutrinos remain an almost
smooth component.  Thus, equations (\ref{eq:halomodel}) to (\ref{eq:sigma}) are
evaluated for $P^{\rm L}(k,z) = P^{\rm L}_{b+c}(k,z)$, where $P^{\rm L}_{b+c}(k,z)$ is
the combined baryon and CDM linear power spectrum defined in (\ref{eq:pbc}).  Correspondingly,
we take the average matter density  in halos to be
$\bar{\rho}_{\rm halo} \simeq (\Omega_b+\Omega_c) \ \rho_{\rm crit} $.
The corrected spectrum $P^{\rm NL}_{b+c}(k,z)$ is then added with the linear neutrino
spectrum $P^{\rm L}_\nu(k,z)$ to form the total matter power spectrum $P_m(k,z)$ as per
(\ref{eq:pm}).

\end{document}